\documentclass[preprint]{elsarticle}
\usepackage[utf8]{inputenc}
\usepackage[margin=3cm]{geometry}
\biboptions{numbers,sort&compress}
\usepackage{csquotes}\MakeOuterQuote{"}
\usepackage{cite}
\usepackage{verbatim,amsmath,amssymb,amsfonts,float,hyperref,parskip,bm}
\usepackage[dvipsnames]{xcolor}
\usepackage{sectsty, soul}

\sectionfont{\bfseries}
\subsectionfont{\bfseries}
\DeclareGraphicsExtensions{.png,.pdf}

\title{Signaling receptor localization maximizes cellular information acquisition in spatially-structured, natural environments}

\author[1]{Zitong Jerry Wang}
\ead{zwang2@caltech.edu}
\author[1]{Matt Thomson}
\ead{mthomson@caltech.edu}
\address[1]{Division of Biology and Biological Engineering, California Institute of Technology, Pasadena, California, 91125, USA}

\date{March 2021}

\begin{document}

	

	
	\begin{abstract}
	Cells in natural environments like tissue or soil sense and respond to extracellular ligands with intricately structured and non-monotonic spatial distributions that are sculpted by processes such as fluid flow and substrate adhesion. Nevertheless, traditional approaches to studying cell sensing assume signals are either uniform or monotonic, neglecting spatial structures of natural environments. In this work, we show that spatial sensing and navigation can be optimized by adapting the spatial organization of signaling pathways to the spatial structure of the environment. By viewing cell surface receptors as a sensor network, we develop an information theoretic framework for computing the optimal spatial organization of a sensing system for a given spatial signaling environment. Applying the framework to simulated  environments, we find that spatial receptor localization maximizes information acquisition in many natural contexts, including tissue and soil. Receptor localization extends naturally to produce a dynamic protocol for redistributing signaling receptors during cell navigation and can be implemented in a cell using a feedback scheme. In a simulated tissue environment, dynamic receptor localization boosts navigation efficiency by 30-fold.  Broadly, our framework readily adapts to studying how the spatial organization of signaling components other than receptors can be modulated to improve cellular information processing.
	\end{abstract}
	
	\maketitle
	
	\section*{Introduction}
	
	Cells sense and respond in spatially-structured environments, where signal distributions are determined by a range of chemical and physical processes from substrate adhesion to fluid flow \citep{fowell2021spatio}. In tissue and soil, distributions of extracellular ligands can be discontinuous, consisting of local ligand patches that differ strongly from monotonic gradients \citep{yang2007binding,weber2013interstitial,russo2016intralymphatic,milde2008hybrid,kicheva2007kinetics,sarris2012inflammatory,kennedy2006axon,raynaud2014spatial,hodge2006plastic,de2021chemotaxis,nunan2001quantification,lim2015neutrophil,von2006growth}. In tissue, diffusive signaling molecules are transported by interstitial fluid through a porous medium. These molecules are then captured by cells and a non-uniform network of extracellular matrix (ECM) fibers, taking on a stable, yet uneven distribution \citep{russo2016intralymphatic,weber2013interstitial,milde2008hybrid,kicheva2007kinetics,sarris2012inflammatory,kennedy2006axon}. For example, ECM-bound chemokine (CCL21) gradients extending from lymphatic vessels take on stable spatial structures, characterized by regions of high ligand concentration separated by spatial discontinuities \citep{weber2013interstitial}. Similar observations have been made for the distribution of other chemokines, axon guidance cues and morphogens in tissues \citep{kicheva2007kinetics,sarris2012inflammatory,kennedy2006axon,lim2015neutrophil}. In soil, a heterogeneous pore network influences the spatial distribution of nutrients by dictating both the locations of nutrient sources as well as where nutrients likely accumulate \citep{raynaud2014spatial,hodge2006plastic,de2021chemotaxis,nunan2001quantification}. Free-living cells detect chemical cues released by patchy distributions of microorganisms, where molecules are moved via fluid flow and diffusion \citep{raynaud2014spatial,hodge2006plastic}. Cells in these and other natural environments experience surface ligand profiles with varying concentration peaks, non-continuity and large dynamic range \citep{kennedy2006axon, dlamini2020combinatorial}, far different from that of a smoothly varying, purely-diffusive environment.

Modern signal processing theory shows that sensing strategies must adapt to the statistics of the input signals, suggesting that spatial sensing in cells should be adapted to the spatial structure of signaling molecules in the cells' native environments \citep{candes2008introduction}. For example, when designing electronic sensor networks sensing spatial phenomena, adapting sensor placement to the spatial statistic of the signal can significant improve information acquisition. Similar considerations may apply to the spatial organization of cell signaling pathways. Furthermore, spatial navigation where sensing plays a key role may also benefit from such adaptation, as suggested by work from robot navigation \citep{iida2016adaptation}. For example, cells navigating up interstitial gradients can potentially get trapped by local concentration peaks \citep{weber2013interstitial}. Adapting sensing to patchy structure of the gradient may allow cells to overcome local traps.

 
	Traditional approaches to studying cell sensing often use highly simplified environmental models, where signals are either uniform or monotonic, neglecting the complex spatial structure in natural cell environments \citep{berg1977physics,hu2010physical,mugler2016limits,endres2008accuracy}. Classic work beginning with the seminal paper by Berg and Purcell (1977) studied cell sensing in homogeneous environments \citep{berg1977physics}. This and subsequent works were extended to study the detection of spatially-varying concentrations, where monotonic gradients remain the canonical environmental model \citep{hu2010physical,mugler2016limits,endres2008accuracy}. Recent work studied more complex sensing environment by adding spatially-uncorrelated fluctuations to a monotonic gradient, which does not capture the spatial structure of natural environments \citep{chou2011noise}. As a practical consequence, little effort in cell engineering has gone into addressing challenges posed by non-monotonic spatial distribution of ligands found in natural environments \citep{martinez2019car}. Fundamentally, it’s not clear what sense and response strategies are well-adapted to operate in environments where signals are distributed in complex spatial patterns.

	Observations of dynamic receptor rearrangement in leukocytes, neurons, and keratinocytes suggest that cells might modulate the placement of surface receptors to exploit the spatial structure of ligand distribution \citep{pignata2019spatiotemporal,bouzigues2007asymmetric,nieto1997polarization,van2003leukocyte,shimonaka2003rap1,yokosuka2005newly,mossman2005altered,fang1999epidermal}. For example, multiple classes of axon guidance receptors can dynamically rearrange on the surface of growth cones \citep{pignata2019spatiotemporal,bouzigues2007asymmetric}. In all such cases, receptors rearrange constantly, adjusting local surface densities in response to changes in ligand distribution across the cell surface. Chemokine receptors in lymphocytes and growth factor receptor in keratinocytes exhibit similar spatial dynamics \citep{nieto1997polarization,van2003leukocyte,shimonaka2003rap1, fang1999epidermal}. However, there are also T cell and neutrophil receptors that are constantly uniform, even when ligands are distributed non-uniformly \citep{vicente2004role}.
	In addition, during antigen recognition, T-cell receptors (TCRs) take on different placements, ranging from uniform to highly polarized, depending on the density of antigen molecules on the surface of the opposing cell \citep{majzner2020tuning}. 
Thus, across a diverse range of cell surface receptors, we see different,
even contradictory rearrangement behavior in response to changes in environmental structure. It remains unclear whether dynamic receptor rearrangement has an overarching biological function across disparate biological contexts.
	
	We formulate a mathematical framework to solve for receptor placements that maximizes information acquisition in natural environments, generating such environments using existing computational models of tissue and soil microenvironments. Using this framework, we show that dynamic localization of receptors is an effective spatial sensing strategy in natural cell environments, but inconsequential in purely diffusive environments. Thus, anisotropic receptor dynamics previously observed in cells are nearly optimal. Specifically, information acquisition is maximized when receptors are localized and oriented, forming a cap at the region of highest ligand concentration. This placement strategy offers significant improvement over uniformly distributed receptors, but only in natural environments, leading to 2 fold increase in information acquisition. Receptor localization maximizes information acquisition by taking advantage of patchy ligand distribution, reallocating sensing resource from low signal region to a small but high signal region on the cell surface where most of the information is concentrated. 
	
Our framework extends naturally to produce a dynamic protocol for continuously relocalizing receptors in response to a dynamic environment. We show that a simple feedback scheme implements this protocol within a cell, and improves cell navigation significantly. Compared to cells with uniform receptor placement, cells using this scheme achieve more than 30-fold improvement in their ability to localize to the peak of simulated interstitial gradients. Since this strategy is purely spatial, it can be applied across a wide range of chemical environments. Taken together, our model serves as a useful conceptual framework for understanding the role of spatial organization of signal transduction pathway in spatial sensing, and provides a sensing strategy that is both effective in natural cell environments and amenable to cell engineering.
	
	
	\section*{Results}
	\subsection*{An optimal coding framework allows the computation of optimal receptor placement given spatial signal statistics}
	
	\begin{figure}[H]
		\includegraphics[width=14cm]{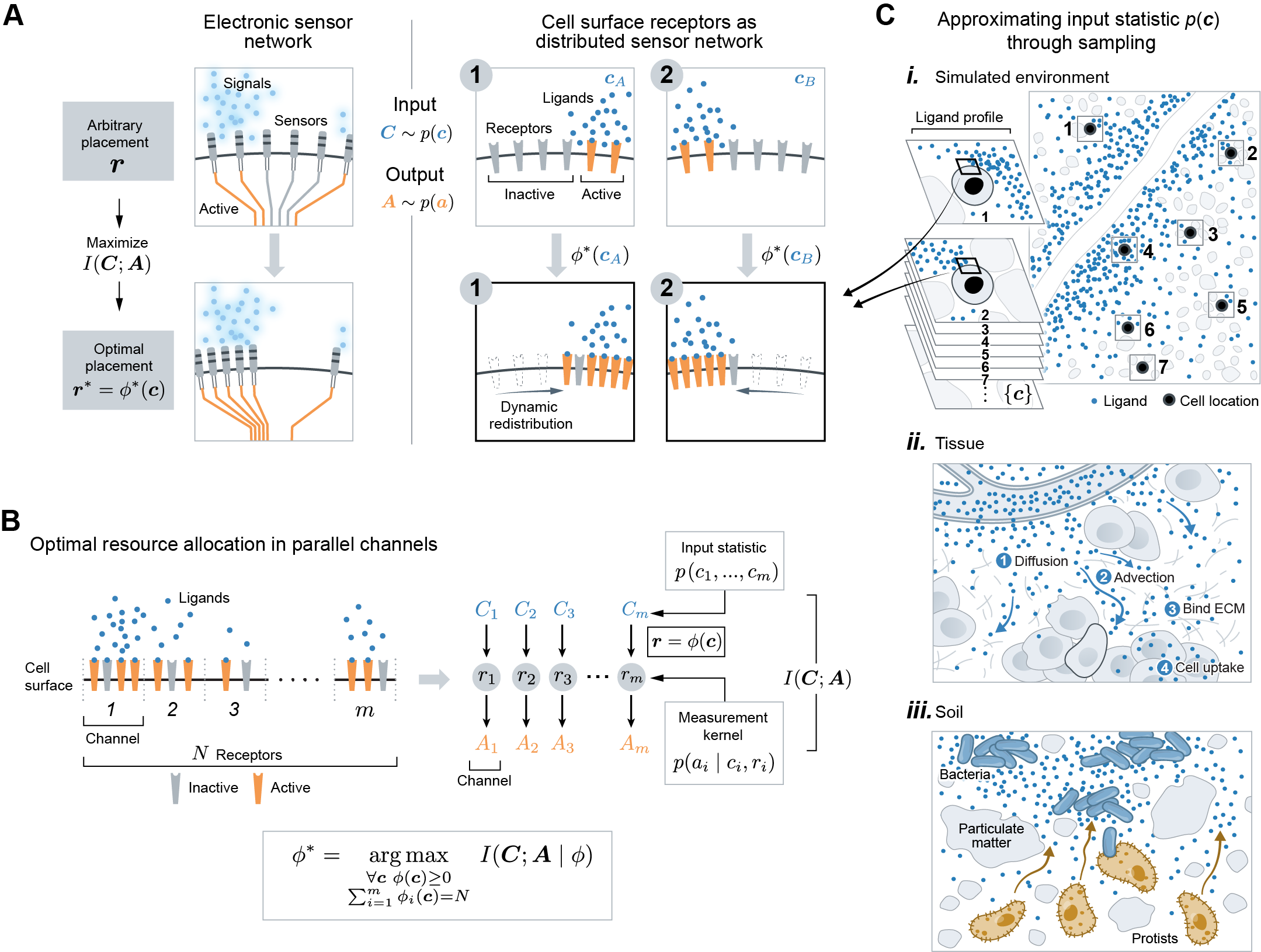}
		\centering
		\caption{\textbf{Adapting receptor placements to signal (input) statistic of natural cell environments} (A) (Left) tuning sensor placement significantly boosts the performance of electronic sensor network. (Right) cell surface receptors also function as a sensor network, taking as inputs ligand profiles $\bm{C}$ across the cell surface and producing as outputs a profile of receptor activity $\bm{A}$ across the cell membrane. The optimal receptor placement strategy $\phi^*: \bm{c} \rightarrow \bm{r}$ maps each ligand profile to a receptor placement, such that the mutual information $I(\bm{C};\bm{A})$ is maximized; $m=100$. (B) The problem of optimal receptor placement can be formulated as a resource allocation problem over parallel, noisy communication channels. The $i$-th channel represents the $i$-th region of the cell membrane, with input $C_i$, output $A_i$ and receptor number $r_i$. We solve for the strategy $\phi^*$, which optimally allocates $N$ receptors to $m$ channels for each ligand profile $\bm{c}$. The input statistic $p(\bm{c})$ depends on the environment, and the measurement kernel $p(a_i|c_i,r_i)$ is modeled as a Poisson counting process (C) i. Approximating input statistic by simulating natural environments and sampling ligand profiles $\{\bm{c}\}$ by tiling cells across the environment; ii. modeling ligand distribution in tissue microenvironment by incorporating diffusion, advection, ECM binding, degradation, and cell uptakes.  iii. modeling ligand distribution in soil microenvironment by generating bacteria distributed in spatial patches, releasing diffusive ligands.}
		\label{fig:optsetup}
	\end{figure}
	We are interested in optimal strategies for a task we refer to as spatial sensing. Spatial sensing is an inference task where a cell infers external profiles of varying ligand level across its surface from an internal profile of varying receptor activity across its membrane.
	This is a useful model task since optimizing performance on this task should improve the cell's ability to infer diverse environmental features.
	
	We developed a theoretical framework to study whether manipulating the placement of cell surface receptors can improve the spatial sensing performance. 
	Optimizing spatial sensing by tuning receptor placement is analogous to optimizing distributed electronic sensor network by adjusting the location of sensors, which has been extensively studied in signal processing.
	In the optimization of distributed sensor networks monitoring spatial phenomena (\autoref{fig:optsetup}A), it is well-known that adjusting the placement of a limited number of sensors can significantly boost sensing performance, where the optimal placement strategy is dictated by the statistics of the input signals \citep{krause2008near}. 
	The collection of a limited number of receptors on the cell surface also functions as a (distributed) sensor network, sensing a spatial profile of varying ligand concentration (\autoref{fig:optsetup}A). Therefore, we hypothesized that receptor placement can be tuned to improve spatial sensing, and that the optimal strategy depends on the statistics of ligand profiles that cells typically encounter. 
	Traditionally, sensor network optimization focuses on finding a single placement strategy. However, cells can rearrange their receptors within a matter of minutes \citep{bouzigues2007asymmetric}, thus leading to a potentially much richer class of strategies. Thus, instead of considering a single placement strategy, we examined a function $\phi$ that assigns a receptor placement to each ligand profile.
	We define the optimal placement strategy as the one that maximizes mutual information between ligand profiles and active receptor profiles, while keeping total receptor number fixed. Mutual information quantifies the extent to which observing one random variable (i.e. the membrane profile of active receptors), reduces uncertainties about another (i.e. the surface profile of ligand counts). This metric sets a theoretical bound on the accuracy of spatial sensing. Notably, this metric is agnostic to the decoding process in that it does not assume any details about downstream signaling, nor the exact environmental features a cell may try to decode, expanding the scope of our results.
	
	Before presenting the general optimization problem, we set up the mathematical framework through the lens of information theory. Consider a 2D cell with a 1D membrane surface. By discretizing the membrane into $m$ equally-sized regions, we modeled the membrane-receptor system as $m$ parallel communication channels (\autoref{fig:optsetup}B). The problem can then be formulated as a resource allocation problem in parallel channels. Specifically, we asked how the cell should allocate $N$ receptors among the $m$ channels to maximize "information" between the channels' inputs and outputs. The $i$-th channel takes as input $C_i \in \mathbb{Z}_{\geq 0}$, a random variable denoting ligand count at the $i$-th region of the membrane surface. Given $r_i \in \mathbb{Z}_{\geq 0}$ receptors, this channel produces as output $A_i \in \mathbb{Z}_{\geq 0}$, a random number of active receptors. This randomness is due to both the randomness in $C_i$ and the stochastic nature of receptor activation itself. For all $m$ channels, our model comprised four key mathematical objects: (1) the ligand profile $\bm{C} = (C_1, ..., C_m)$, (2) receptor placement $\bm{r}= (r_1,..,r_m)$, (3) active receptor profile $\bm{A} = (A_1, ..., A_m)$, and (4) measurement kernel $P(\bm{A}=\bm{a} \mid \bm{C}=\bm{c}, \bm{r})$. 
	The input $\bm{C}  \sim p(\bm{c})$ is now the entire ligand profile across the cell surface. Each realization $\bm{c}$ of $\bm{C}$ has probability $p(\bm{c})$ of being observed. The input statistic $p(\bm{c})$ is central to this work. We explain below how $p(\bm{c})$ can be constructed to represent statistics of ligand profiles cells naturally encounter (see \textbf{Input statistic}). The receptor  profile $\bm{r}$ denotes the number of receptor allocated to each membrane region.
	The output  $\bm{A} \sim p(\bm{a})$ is the number of active receptors across the membrane, which depends on $\bm{c}$ and $\bm{r}$ through $p(\bm{a}|\bm{c}, \bm{r})$, the measurement kernel. We explain below how this kernel can be modeled (see \textbf{Measurement kernel}). 
	
	In our general optimization problem (\autoref{fig:optsetup}B), we considered $\bm{r}$ as a parameter that cells can adjust in response to their environment. Thus, our decision variable is the placement strategy $\phi: \bm{c} \rightarrow \bm{r}$, mapping a ligand profile to a receptor placement. We are interested in the choice of $\phi$ that maximizes the amount of "information" the cell can obtain regarding $\bm{C}$ by observing $\bm{A}$, for a fixed number of receptors $N$. Formally, we quantified this information using the mutual information, 
	\begin{align}
		I(\bm{C}; \bm{A}) = \sum_{\bm{c} \in \bm{C}} \sum_{\bm{a} \in \bm{A}} p(\bm{c},\bm{a}) \log \frac{p(\bm{c},\bm{a})}{p(\bm{c})p(\bm{a}) }.
		\label{mutualinfo1}
	\end{align}
	The mutual information is minimized when $\bm{C}$ and $\bm{A}$ are independent, and maximized when one is a deterministic function of the other. Since $p(\bm{c},\bm{a}) = p(\bm{a}|\bm{c}, \bm{r} = \phi(\bm{c}))p(\bm{c})$, each summand in the mutual information will be affected by the choice of $\phi$. Taken together, we arrive at our general formulation of the optimal strategy $\phi^*$:
	\begin{equation}\label{appendixopttrue}
		\phi^* =  \underset{\substack{\forall \bm{c} \,\,  \phi(\bm{c}) \geq 0 \\ \sum_i \phi_i(\bm{c}) = N}}{\text{argmax}} \hspace{4pt} I(\bm{C};\bm{A} \mid \phi),
	\end{equation}
	where $N$ is the total number of receptors. To solve for $\phi^*$, we needed to specify both a measurement kernel $p(a| c,r)$ and an input statistic $p(\bm{c})$:
	
	\textbf{Measurement kernel.} We modeled $p(\bm{a}|\bm{c},\bm{r})$ assuming that each receptor binds ligands locally and activates independently of other receptors. These assumptions allowed $p(a|c,r)$ to factorize as follows,
	\begin{equation}\label{appendixchannel1}
		\begin{split}
			P(\bm{A} = \bm{a} \mid \bm{C} = \bm{c}, \, 
			\bm{r}) &= \prod_{i=1}^m P(A_i=a_i \mid C_i = c_i,\ r_i).\\
		\end{split}
	\end{equation}
	Furthermore, each local sensing process was modeled as a Poisson counting process, such that
	\begin{equation}\label{appendixchannel2}
		\begin{split}
			P(A_i=a_i \mid C_i = c_i,\, r_i) &=  \frac{\mu_i{^{a_i}}}{a_i!}e^{-\mu_i}, \hspace{4pt} \text{where} \,\, \mu_i = r_i \Big(\frac{c_i}{c_i+k_D} + \alpha \Big). \\
		\end{split}
	\end{equation}
	$K_d$ is the equilibrium dissociation constant and $\alpha$ represents constitutive receptor activity, which we take to be small ($\alpha \ll 1$) \citep{slack2012development}. In other words, the number of active receptors $A_i$ given ligand count $c_i$ is a Poisson random variable with mean $\mu_i = r_i(\frac{c_i}{c_i+k_D}+\alpha)$.  Equation \eqref{appendixchannel1} and \eqref{appendixchannel2} together specify the measurement kernel.
	
	\textbf{Input statistic.} Next, we specified the input statistic $p(\bm{c})$ which depends on the class of environment. In this work, we studied three classes of environments: soil, tissue, and monotonic gradient.
	A formal construction of input statistic $p(\bm{c})$ is feasible for simplified environments, but will likely not be analytically tractable for complex ligand profiles from natural environments. Therefore, we took an empirical approach, computationally generating instances of each environment and directly sampled ligand profiles from them (\autoref{fig:optsetup}C, i).
	For each class, we generated ligand environments by simulating a corresponding partial-differential equation (PDE) over a spatial domain $\Omega \subset \mathbb{R}^2$ (see Methods). 
	For tissue, the PDE model incorporated \emph{in vivo} processes such as fluid flow, non-uniform ECM binding and cell uptake, to produce an immobilized interstitial gradient from a localized source (\autoref{fig:optsetup}C-ii, \autoref{fig:efficacy}A). 
	For soil, diffusive ligands are released from a collection of soil bacteria whose spatial distribution was generated from a statistical model (\autoref{fig:optsetup}C-iii, \autoref{fig:efficacy}A). 
	We also considered a monotonic gradient which is an exponential fit to the interstitial gradient of the tissue environment. This fitting ensures any difference between the two environments are due to differences in local structures, not global features such as gradient decay length or average concentration.
	For each environment we generated, we tiled it with a cell of fixed size and evaluated the PDE solution along each cell membrane to obtain a set of ligand profiles denoted $\{\bm{c}\}$ (\autoref{fig:optsetup}C, i). Putting the empirical measure on the samples $\{\bm{c}\}$ approximates the true distribution of $\bm{C}$. 
	It is important to note that although we modeled $p(c)$ and $p(a|c)$ in these ways, the overall framework can accommodate any alternative choices of model.
	
	For these choices of $p(\bm{c})$ and $p(\bm{a}|\bm{c})$, we aimed to study the functional relationship between ligand profiles $\{\bm{c}\}$ and their optimal receptor placements $\phi^*(\bm{c})$. To this end, we optimized receptor profiles for each sampled profile $\bm{c}$ individually, reducing the general problem to a local formulation. Given ligand profile $\bm{c}$, random vector $\hat{\bm{c}}$ represents local fluctuations of $\bm{c}$ due to stochasticity of reaction-diffusion events. In the case of unimolecular reaction-diffusion processes, it can be shown that $\hat{\bm{c}}$ is a Poisson vector with mean equal to $\bm{c}$, solution of the PDE. Therefore, we can solve for $\phi^*(\bm{c})$ locally by maximizing the mutual information between $\hat{\bm{c}}$ and the resulting output $\hat{\bm{a}}$,
	\begin{equation}
		\phi^*(\bm{c})	=  \underset{\substack{  \bm{r} \geq 0 \\ \sum_i r_i = N}}{\text{argmax}} \hspace{4pt}  I(\hat{\bm{c}}, \hat{\bm{a}}\mid \bm{r}),
		\label{eq:localopt}
	\end{equation}
	where $p(\hat{\bm{a}}) = \sum_{\bm{c}} p(\hat{\bm{a}}|\hat{\bm{c}}=\bm{c})p(\hat{\bm{c}}=\bm{c})$ and $N = 10^4$ is the total receptor number. We assume $\bm{r}$ to be real-valued when solving \eqref{eq:localopt}, this is reasonable as long as $N$ is not too small. Note that the resulting set of locally optimal receptor profiles $\{\phi^*(\bm{r})\}$ obtained this way still depends crucially on the input statistics $p(\bm{c})$. Namely, sampled ligand profiles $\{\bm{c}\}$ will be different for classes of environments with different $p(\bm{c})$, hence affecting the kind of receptor profiles in $\{\phi^*(\bm{r})\}$. In this way, the statistical structure over the space of ligand profiles plays an important role in determining which receptor placement is effective, even when the placements are computed locally for each ligand profile.

	\subsection*{Receptor polarization yields optimal spatial sensing in natural environments}
	
	\begin{figure}[H]
		\centering
		\includegraphics[width=15cm]{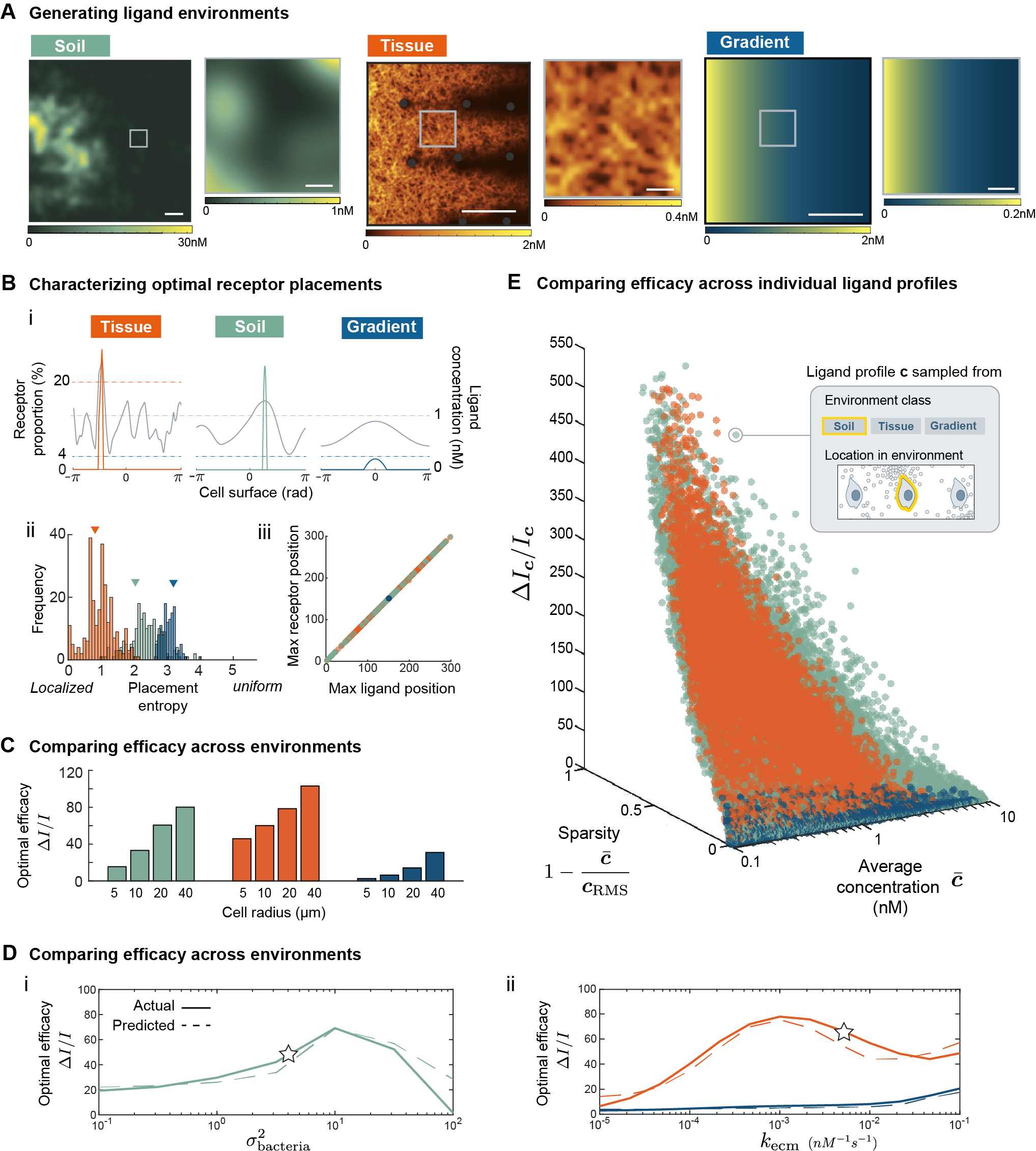}
		\caption{\textbf{Receptor polarization optimizes information acquisition in natural environments.}
			(A), computationally generated ligand fields using PDE models of soil, interstitial gradient (gray dots represent cells), and simple (exponential) gradient, all scalebar $= 100\mu m$.
			(B), i) Example of optimal receptor profile $\phi^*(\bm{c})$ (colored) and the corresponding ligand profile $\bm{c}$ (gray); ii) entropy for each optimal receptor placements in $\{\phi^*(\bm{c})\}$ colored by environment, colored triangles indicate the entropy of three receptor placements shown in i); iii) scatterplot where each dot corresponds to an optimal placement $\phi^*(\bm{c})$, x-axis is membrane position with the most receptor, y-axis is membrane position with most ligand in $\bm{c}$. 
			(C), optimal efficacy $\Delta I/I$ colored by environments, computed with ligand profiles $\{\bm{c}\}$ sampled using cells of different radius.
			(D), i) actual (solid) and predicted (dashed) $\Delta I/I$ for soils with varying values of $\sigma^2_{\text{bacteria}}$, ii) actual and predicted $\Delta I/I$ for tissue with varying values of $k_{\text{ecm}}$, and for exponential gradients fitted to each tissue (gradient). Stars correspond to parameter values used to generate panel A-C and E.
			(E), scatterplot where each dot corresponds to a single pair of $\bm{c}$ and $\phi^*(\bm{c})$: axis are placement efficacy $\Delta I_{\bm{c}}/I_{\bm{c}}$,  average concentration and sparsity of $\bm{c}$; across all panels, $N = 1000$, $k_D = 35 nM$, $\alpha = 0.1$, $m=100$.}
		\label{fig:efficacy}
	\end{figure}
	
	Optimal strategies of receptor placement are similar for soil and tissue environment, where receptors localize into a single cap and orient towards the region of highest ligand concentration.
	\autoref{fig:efficacy}B-i shows three examples of optimal receptor placements $\phi^*(\bm{c})$ (colored) and the corresponding ligand profile $\bm{c}$, one from each class of environments shown in \autoref{fig:efficacy}A.
	Compared to monotonic gradient, receptors are much more localized for the ligand profiles sample from tissue and soil, nearly all of which are found within $1\%$ of the entire membrane.
	In all three cases, the peak of each optimal receptor profile (colored) is oriented towards the position of highest ligand concentration.
	Indeed, both features of localization and orientation holds in general.
	Firstly, across all sampled ligand profiles $\{\bm{c}\}$, \autoref{fig:efficacy}B-ii shows that optimal receptor profiles tend to have low entropy. 
	The entropy of receptor profile $\bm{r}$, defined as $\sum_i r_i \log r_i$, can be used as a measure of localization. 
	Low entropy corresponds to profiles where most receptors are concentrated within a few membrane positions. 
	Such strong localization (low entropy) is partly due to the nonlinearity of the (Poisson) channel mutual information with respect to the ligand concentration (see Discussion), and is more prevalent in soil and tissue due to their larger local variations in ligand concentration compared to monotonic gradient (\autoref{fig:efficacy}A).
	Additionally, the optimal strategy consistently allocated the maximum number of receptors to the position of maximum ligand concentration (\autoref{fig:efficacy}B-iii).
	Although not common, it is possible for the optimal placement to consist of multiple receptor peaks, which occurs when there are multiple ligand peaks of very similar concentration.
	Note that ligand concentration also influences the optimal strategy, which we take to be highly dilute in agreement with empirical measurements (see Methods). In saturating environments, however, the optimal solution can be qualitatively different (see SI).
	In summary, the optimal placement strategy $\phi^*$ in all environments studied can be well-approximated by a simple winner-takes-all scheme, where receptors polarize to form a single cap at the position of maximum ligand concentration.
	
	Optimally placed receptors significantly improve information acquisition relative to uniform receptors, but only in soil and tissue environments.
	To make this statement precise, we quantified the efficacy of a placement strategy $\phi: \bm{c} \rightarrow \bm{r}$ with respect to a set of ligand profiles $\{\bm{c}\}$. The efficacy of $\phi$ is the relative increase in average information cells acquire by adapting the strategy $\phi$ compared to a uniform strategy $\phi^u$, 
	\begin{equation}\label{eq:deltaII}
		\Delta I/I(\phi) =  \frac{\langle I(\hat{\bm{c}};\hat{\bm{a}} \mid \phi)\rangle_{\bm{c}} - \langle I(\hat{\bm{c}};\hat{\bm{a}} \mid \phi^u)\rangle_{\bm{c}}}{\langle I(\hat{\bm{c}};\hat{\bm{a}} \mid \phi^u)\rangle_{\bm{c}}},
	\end{equation}
	where $\langle \cdot\rangle$ denotes averaging across the set of sample ligand profiles $\{\bm{c}\}$, and recall $\hat{\bm{c}}$ is a Poisson vector with mean $\bm{c}$. We are specifically interested in the optimal efficacy $\Delta I/I(\phi^*)$, and simply refer to it as $\Delta I/I$ when the dependency is clear from context.
	For $\Delta I/I(\phi^*)$, the set of ligand profiles $\{\bm{c}\}$ contained in its definition is always the same set that $\phi^*$ is optimized for.
	The larger $\Delta I/I$ is, the more beneficial it is for cells to place receptors optimally rather than uniformly. 
	We found that $\Delta I/I$ is an order-of-magnitude larger for soil and tissue environment compared to an exponential gradient (\autoref{fig:efficacy}C). In other words, placing receptors optimally rather than uniformly benefits cells in complex, natural environments more than cells in monotonic gradients. Furthermore, this difference between environments persist for cells of different size  (\autoref{fig:efficacy}D). Note that differences between tissue and monotonic gradient are due to differences in local spatial structure, not global features such as gradient decay length or global average concentration, as both parameters were made to be identical between the two environments.
	
	For both soil and tissue environment, the optimal efficacy depends on a key parameter in their respective PDE model. We illustrate this dependence by adjusting the value of each respective parameter, sample new ligand profiles $\{\bm{c}\}$, solve for optimal placements $\{\phi^*(\bm{c})\}$, and compute optimal efficacy $\Delta I/I$. \autoref{fig:efficacy}D shows how the optimal efficacy (solid lines) changes as we adjust environmental parameters. In soil, $\Delta I/I$ dropped substantially when ligand sources (bacteria) were either uniformly dispersed or highly clustered (\autoref{fig:efficacy}D-i), corresponding to values of the parameter $\sigma^2_{\text{bacteria}}$ much larger or smaller than an empirical estimate (star in \autoref{fig:efficacy}D-i) taken from literature \citep{raynaud2014spatial}, respectively. This result is intuitive since uniformly distributed sources create a nearly homogeneous environment, whereas grouping all sources into a single cluster produces an environment similar in structure to that with only a single point source. In tissue, $\Delta I/I$ dropped when most ligands were found in solution, instead of bound to the ECM (\autoref{fig:efficacy}D-ii), corresponding to low ECM binding rate ($k_{ecm}$). For reference, star indicates the empirical value of $k_{ecm}$ for the chemokine CXCL13 \citep{yang2007binding}.
	Nonetheless, compared to its fitted exponential gradient, $\Delta I/I$ in the interstitial gradient remain significantly higher for all ECM binding rates (\autoref{fig:efficacy}E, ii). In tissue, gradients made up of ECM-bound ligands are ubiquitous, suggesting the optimization of receptor placement is highly relevant in such an environment.
	
	Surprisingly, the differences in optimal efficacy between environments can be well-explained by simple differences between the ligand profiles they generate. Specifically, optimal efficacy is larger in soil and tissue because ligand profiles $\{\bm{c}\}$ sampled from such environments tend to be more "patchy", having most of the ligands concentrated in a small subset of membrane regions. We make this statement precise by defining two measures on $\bm{c}$. First, we quantified patchiness of a ligand profile $\bm{c}$ using a measure of sparsity,
	\begin{equation}
		\text{sparsity}(\bm{c}) = 1 - \frac{\ \overline{\bm{c}}\ }{\ \bm{c}_{\text{RMS}}\ },
	\end{equation}
	where $\bm{c}_{\text{RMS}}$ is the root-mean-square and $\overline{\bm{c}} = \frac{1}{m} \sum_i c_i$ is the average ligand concentration. A ligand profile with a sparsity of one has all ligands contained in a single membrane region, whereas a uniform distribution of ligands has a sparsity of zero. Second, we defined an efficacy measure for each ligand profile $\bm{c}$,
	\begin{equation}\label{eq:deltaII}
		\Delta I_{\bm{c}}/I_{\bm{c}} =  \frac{ I(\hat{\bm{c}};\hat{\bm{a}} \mid \phi^*) - I(\hat{\bm{c}};\hat{\bm{a}} \mid \phi^u)}{I(\hat{\bm{c}};\hat{\bm{a}} \mid \phi^u)},
	\end{equation}
	Analogous to $\Delta I/I$, $\Delta I_{\bm{c}}/I_{\bm{c}}$ measures relative increase in information for a particular ligand profile $\bm{c}$ instead of averaging across the entire set $\{\bm{c}\}$. The larger $\Delta I_{\bm{c}}/I_{\bm{c}}$ is, the more useful the optimal placement is for sensing $\bm{c}$ compared to a uniform placement. We computed $\Delta I_{\bm{c}}/I_{\bm{c}}$ and sparsity for each ligand profile sampled, across all environments. \autoref{fig:efficacy}E shows that 1) sparser ligand profiles tend to have higher $\Delta I_c/I_c$, and 2) ligand profiles sampled from soil and tissue tend to be sparser compared to profiles from monotonic gradients. Thus optimizing receptor placement may be more effective for sensing in natural environments because such environments generate sparser ligand profiles. \autoref{fig:efficacy}E also shows a weaker, negative relationship between $\Delta I_{\bm{c}}/I_{\bm{c}}$ and average concentration. 
	In fact, using just these two measures, sparsity and average concentration of each ligand profile $\bm{c}$, we can predict the optimal efficacy $\Delta I/I$ over the entire set $\{\bm{c}\}$. 
	To illustrate, we fit a linear regression model of $\Delta I/I$ from \autoref{fig:efficacy}C (see SI) using only the sparsity and average concentration of ligand profiles in $\{\bm{c}\}$. \autoref{fig:efficacy}D shows that this simple model accurately predicts $\Delta I/I$ (dashed line) across all three classes of environments and for different parameter values. In conclusion, signals (ligand profiles) from natural environments are dilute and sparsely distributed. In such cases, cells can significantly improve their spatial sensing performance by localizing receptors to the membrane region of maximum ligand concentration.

	\subsection*{Spatial sensing via the optimal strategy is robust to imprecise placements caused by biological constraints}
	
	\begin{figure}[H]
		\includegraphics[width=15cm]{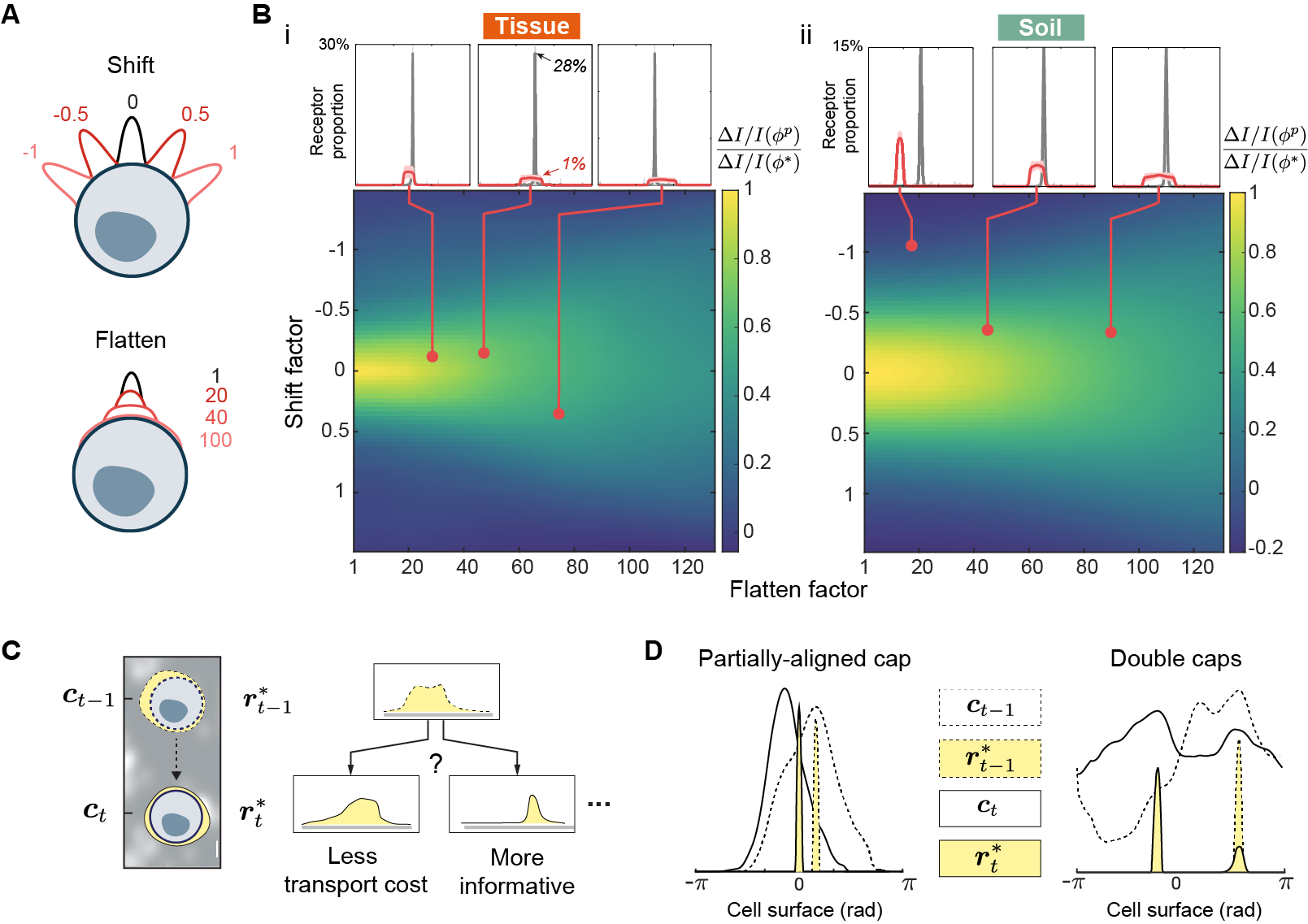}
		\centering
		\caption{\textbf{Optimal efficacy $\Delta I/I (\phi^*)$ is robust to minor deviations in receptor placement away from the optimal form.}
			(A), the effect of different degrees of shifting and flattening applied to a receptor profile (black curve).
			(B), colors of heat map represent ratio of perturbed efficacy $\Delta I/I (\phi^p)$ to optimal efficacy $\Delta I/I(\phi^*)$ for different combinations of shifting and flattening, computed for ligand profiles $\{\bm{c}\}$ sampled from either soil or tissue; call-out boxes corresponds to different sets of perturbations, showing the average of the optimal $\{\phi^*(\bm{c})\}$ (gray) and perturbed $\{\phi^p(\bm{c})\}$ (red) receptor placements, after all profile peaks were centered.
			(C), a moving cell experiences a dynamic ligand profile such that the optimal placement $\bm{r}_t^*$ at time $t$ must trade off information gain for lower transport/redistribution cost, representing latency in receptors redistributing from $\bm{r}^*_{t-1}$ to $\bm{r}_t^*$.
			(D), accounting for transport cost, the optimal placement strategy is modified to polarize receptors at an intermediate position between subsequent ligand peaks or form multiple receptor caps}
		\label{fig:robustness}
	\end{figure}
	
	
	Despite the optimal strategy $\phi^*$ being strongly polarized and precisely oriented, we found that neither features are necessary to achieve near-optimal efficacy. 
	Before presenting the result, we emphasize that this robustness is crucially important as it makes the strategy feasible in cells, since it is likely difficult for real receptors to adopt $\phi^*$ precisely.
	Such difficulty is due to two aspects of $\phi^*$: 1) strong polarization, 2) precise peak-to-peak alignment. 
	Firstly, polarization of membrane proteins, although commonly observed in cells, are far weaker than what the optimal requires. There are also physical limits to how densely packed receptors can be, preventing them from being placed optimally even if they can be controlled precisely.
	Secondly, stochastic fluctuations in ligand profile makes it difficult to precisely align receptor peak to ligand peak.
	Fortunately, receptor placements that are significantly less polarized and misaligned can still achieve near-optimal efficacy.
	To illustrate, we perturb optimal placements and show that sensing efficacy persists when receptors align only partially with ligand peak and polarize weakly. 
	For soil and tissue, we circularly shift and flatten all optimal receptor profiles $\{\phi^*(\bm{c})\}$ computed from sample ligand profiles to obtain $\{\phi^p(\bm{c})\}$, the corresponding set of perturbed profiles.
	\autoref{fig:robustness}A shows the result of applying different degrees of shifting and flattening to a receptor profile (black). 
	Different degrees of shifting and flattening represents different degrees of misalignment and weakened polarization, respectively.
	We want to assess the effect of these perturbations on sensing. 
	To do so, we compute the perturbed efficacy $\Delta I/I(\phi^p)$ using the perturbed profiles, and compare it to the optimal efficacy $\Delta I/I(\phi^*)$.
	The phase diagrams in \autoref{fig:robustness}B shows the ratio of perturbed to optimal efficacy for various combinations of perturbations, across soil and tissue.
	\autoref{fig:robustness}B-i shows examples of perturbations (red dots) that drastically alter the receptor profile while still achieving near-optimal efficacy.
	The red and gray curve in the call-out box represents what the "average" perturbed and optimal profiles look like, respectively. They are obtained by centering the peak of all profiles in $\{\phi^p(\bm{c})\}$ and $\{\phi^*(\bm{c})\}$ followed by averaging across each set element-wise.
	Note that only a modest enrichment of receptors ($1\% \approx 3\times$ relative to uniform) is sufficient to achieve near-optimal efficacy, such degree of polarization has been observed for membrane receptors \citep{bouzigues2007asymmetric, mcclure2015role}.
	Lastly, the phase diagrams in \autoref{fig:robustness}B show that weakly polarized receptors (larger flatten factor) are more robust to misalignment, as indicated by small change in efficacy with large shift factor.
	These results suggest that receptor polarization that is biologically plausible is effective for spatial sensing.
	
	\subsection*{Optimization framework extends naturally to produce a dynamic protocol for sensing time-varying ligand profiles}
	
	Our framework extends naturally to produce a dynamic protocol for rearranging receptors in response to dynamically changing ligand profiles.
	So far, we have viewed ligand profiles as static snapshots and considered instantaneous protocols for receptor placement. 
	In reality, cells sense while actively exploring their environment, so that the ligand profile it encounters is changing in time, both due to intrinsic changes in the environment state as well as due to the motion of the cell.
	As the signal profile $\bm{c}(t)$ changes, we want receptors to redistribute in an "efficient" manner for information acquisition.
	Specifically, we obtain a dynamic protocol by extending our framework to account for both information acquisition and a “cost” for changing receptor location (\autoref{fig:robustness}C). 
	We quantify this cost using the Wasserstein distance $W_1(\bm{r}_A,\bm{r}_B)$, which is the minimum distance receptors must move across the cell surface to redistribute from $\bm{r}_A$ to $\bm{r}_B$. This distance assumes receptors are moved in an "optimal" manner. 
	For a cell sensing a sequence of ligand profiles $\{\bm{c}_t\}_t$ over time, the optimal receptor placement $ \bm{r}^*_{t}$ for $\bm{c}_t$ now depends additionally on $\bm{r}^*_{t-1}$, the optimal placement for the previous ligand profile,
	\begin{equation}
		\bm{r}^*_{t}= \underset{\bm{r} \geq 0 \atop \sum_i r_i = N}{\operatorname{argmax}} \,\, I\left(\hat{\bm{c}}_t; \hat{\bm{a}} \mid \bm{r} \right)-\gamma W_{1}\left( \bm{r}_{t-1}^*, \bm{r} \right),
		\label{eq:dynamicopt}
	\end{equation}
	where $p(\hat{\bm{a}}) = \sum_{\bm{c}} p(\bm{a}|\hat{\bm{c}}_t)p(\hat{\bm{c}}_t)$, and $\gamma \geq 0$ represents the cost of moving one receptor per unit distance. 
	This dynamic formulation admits a natural interpretation as maximizing information rate (information per receptor-distance moved) instead of absolute information gain.
	For $t=1$, we define $r^*_t$ according to the original (static) formulation of equation $\eqref{eq:localopt}$. Hence, one can view this dynamic protocol as the general optimal strategy as it encompasses $\phi^*$.  
	\autoref{fig:robustness}D illustrates two salient features of this dynamic protocol. When the peak of the previous receptor profile $\bm{r}^*_{t-1}$ is near the peak of the current ligand profile $\bm{c}_t$, $\bm{r}^*_{t}$ is obtained by simply shifting $\bm{r}^*_{t-1}$ towards the current ligand peak but not aligning fully (left). When the peak of the previous receptor profile is far from the current ligand peak, some receptors are moved to form a second cap at the current ligand peak (right).
	
	\subsection*{Simple feedback scheme rearranges receptors to achieve near-optimal information acquisition}
	
	\begin{figure}[H]
		\includegraphics[width=15cm]{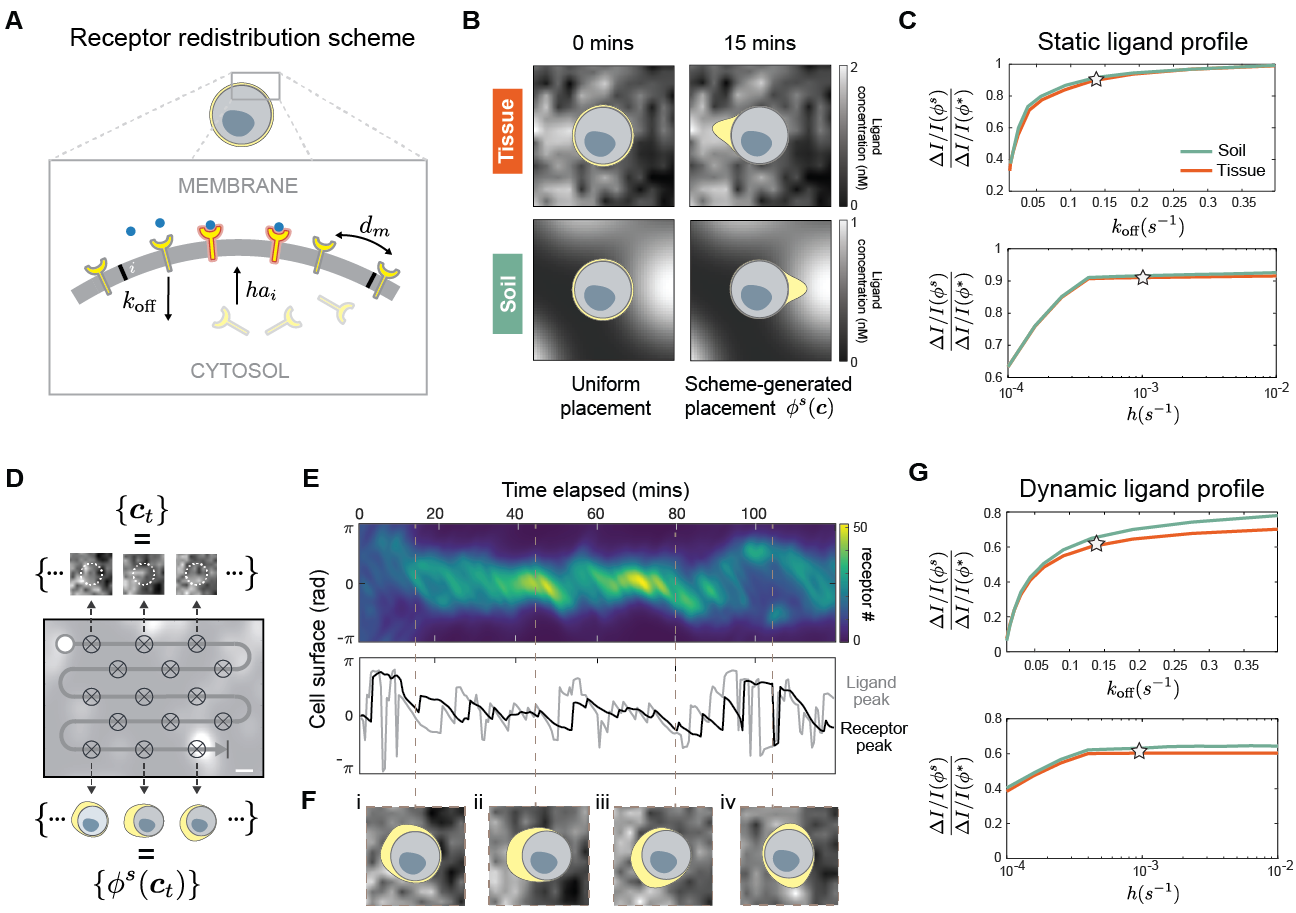}
		\centering
		\caption{
			\textbf{Positive feedback scheme redistributes receptors to achieve near-optimal sensing efficiency for both static and dynamic signals.}
			(A), the cell is modeled as a one-dimensional membrane lattice with a well-mixed cytosol. Receptors are subject to three redistribution mechanisms: receptor endocytosis ($k_{\text{off}}$), receptor activity-dependent transport ($ha_i$), membrane diffusion ($d_m$). 
			(B), receptors profiles (yellow) generated by simulating the feedback scheme for an initially uniform set of receptors, against a static ligand profile from tissue and soil.
			(C), ratio of scheme efficacy $\Delta I/I(\phi^s)$ to optimal efficacy $\Delta I/I(\phi^*)$ for static signals $\{\bm{c}\}$ sampled from soil and tissue, star indicates feedback scheme parameters used for simulation in panel B.
			(D), as a moving cell encounters a sequence of ligand profiles $\{\bm{c}_t\}$, the feedback scheme continuously rearrange receptors, generating a sequence of receptor profiles $\{\phi^s(\bm{c}_t)\}$.
			(E), (top) kymograph showing the entire temporal sequence of receptor profiles of a moving cell; (bottom) position of ligand peak aligned in time with position of receptor peak as generated by the feedback scheme.
			(F), snapshots of receptor profiles taken at select time points.
			(G), ratio of scheme efficacy $\Delta I/I(\phi^s)$ to optimal efficacy $\Delta I/I(\phi^*)$ for a sequence of signals $\{\bm{c}_t\}$ sampled by moving a cell through soil and tissue environment, stars indicate feedback scheme parameters used for simulation in panel E-F.}
		\label{fig:circuit}
	\end{figure}
	
	A positive feedback scheme polarizes and orients receptors to achieve near-optimal information acquisition.
	Polarization is a fundamental building block of many complex spatial behavior in cells, involved in sensing, movement, growth, and division. Many natural polarization circuits are well-characterized down to molecular details. In fact, even synthetic polarization networks have been experimentally constructed in yeast, capable of reliably organizing membrane-bound proteins into a localized pole.
	Such works demonstrate the feasibility of engineering new polarization systems in cells. 
	Using a PDE model of a receptor redistribution scheme, we show that simple, local interactions can redistribute receptor to achieve near-optimal information acquisition, for both static and dynamic signals.
	\autoref{fig:circuit}A illustrates the three redistribution processes (arrows) in our model:
	1) receptors diffuse laterally on the membrane with uniform diffusivity $d_m$,
	2) receptor endocytose with constant rate $k_{\text{off}}$,
	3) receptors are incorporated from a homogeneous cytoplasmic pool to membrane position $i$ with rates $h a_i$, where $a_i$ is the local receptor activity and $h$ a proportionality factor.
	This last process provides the necessary feedback that enables receptor activity to influence receptor placement.
	Budding yeasts achieve this feedback using an interacting loop between intracellular polarity factor Cdc42 and Ste2 receptors \citep{hegemann2015cellular}.
	Note that our feedback scheme is only meant to illustrate one of many possible implementation of the dynamic rearrangement protocol, alternatives such as adjusting receptor endocytosis or lateral mobility are possible, and may be better for cell engineering purposes.
	
	Given a fixed ligand profile $\bm{c}$, \autoref{fig:circuit}B shows our feedback scheme can, within minutes,  localize and orient receptors (yellow region) towards the position of maximum ligand concentration.
	We denote the steady-state receptor profile generated by our scheme as $\phi^s(\bm{c})$.
	As \autoref{fig:circuit}B already shows, scheme-generated profiles are far less polarized than their optimal counterpart $\phi^*(\bm{c})$. Despite this, \autoref{fig:circuit}C shows scheme efficacy $\Delta I/I (\phi^s)$ are close to that of the optimal value $\Delta I/I (\phi^*)$. 
	Recall $\Delta I/I (\phi^*)$ measures the relative increase in average information acquired using optimally-placed instead of uniform receptors. Scheme efficacy $\Delta I/I (\phi^s)$, therefore, makes a similar comparison between scheme-driven and uniform receptors. In \autoref{fig:circuit}C, we see scheme efficacy is robust to variations in both endocytosis ($k_{\text{off}}$) and transport proportionality factor ($h$), when other parameters are fixed to empirical values reported in literature \citep{marco2007endocytosis}. Stars represent values used to obtain profiles in \autoref{fig:circuit}D.
	
	Our feedback scheme can continuous rearrange receptors in response to dynamic ligand profiles, exhibiting dynamics similar to the optimal dynamic protocol.
	\autoref{fig:circuit}D shows that as a cell moves across an environment, it encounters a sequence of varying ligand profiles $\{\bm{c}_t\}$ over time. In response, the scheme redistributes receptors, generating a corresponding sequence of receptor profiles $\{\phi^s(\bm{c}_t)\}$.
	In this dynamic setting, the scheme can still induce asymmetric redistribution of receptors.
	For a cell moving in the tissue environment, \autoref{fig:circuit}E clearly shows this dynamic asymmetry through a kymograph (top) of the entire temporal sequence $\{\phi^s(\bm{c}_t)\}$. Furthermore, snapshots taken along this sequence show receptors polarized towards regions of high ligand concentration as desired (\autoref{fig:circuit}F).
	Interestingly, receptor placements generated by our scheme exhibits both features of the placement strategy derived by accounting for receptor latency, shown in \autoref{fig:robustness}D.
	First, as the ligand peak changes position slightly, the receptor peak gets shifted in the same direction with a delay.
	\autoref{fig:circuit}E illustrates this phenomena by aligning the time trace of both peak positions (bottom). A shift in the ligand peak (gray) is often followed by a corresponding shift in receptor peak (black) after around a $5$-minute delay.
	Second, if the ligand peak changes position drastically, a second receptor cap forms, oriented towards with the new ligand peak.
	\autoref{fig:circuit}F-iv illustrates this phenomena, showing a new receptor cap forming precisely after a large shift in ligand peak position (\autoref{fig:circuit}E).
	We assess the performance of our scheme by comparing scheme-generated placements $\{\phi^s(\bm{c})\}$ and optimal placements $\{\phi^*(\bm{c})\}$ corresponding to the same sequence of ligand profiles $\{\bm{c}_t\}$, across both tissue and soil. 
	\autoref{fig:circuit}G shows that for cells moving in either environment, scheme efficacy $\Delta I/I (\phi^s)$ (star) is not far from the optimal value $\Delta I/I (\phi^*)$. 
	Furthermore, scheme efficacy is robust to variations in endocytosis ($k_{\text{off}}$) and transport proportionality factor ($h$), for all other parameters fixed.
	Taken together, our feedback scheme self-organizes receptors to achieve near-optimal information acquisition, in both static and dynamic environments.
	
	\subsection*{Feedback scheme enables cells to search quickly and localize precisely in simulated interstitial gradients}
	\begin{figure}[H]
	
\includegraphics[width=15cm]{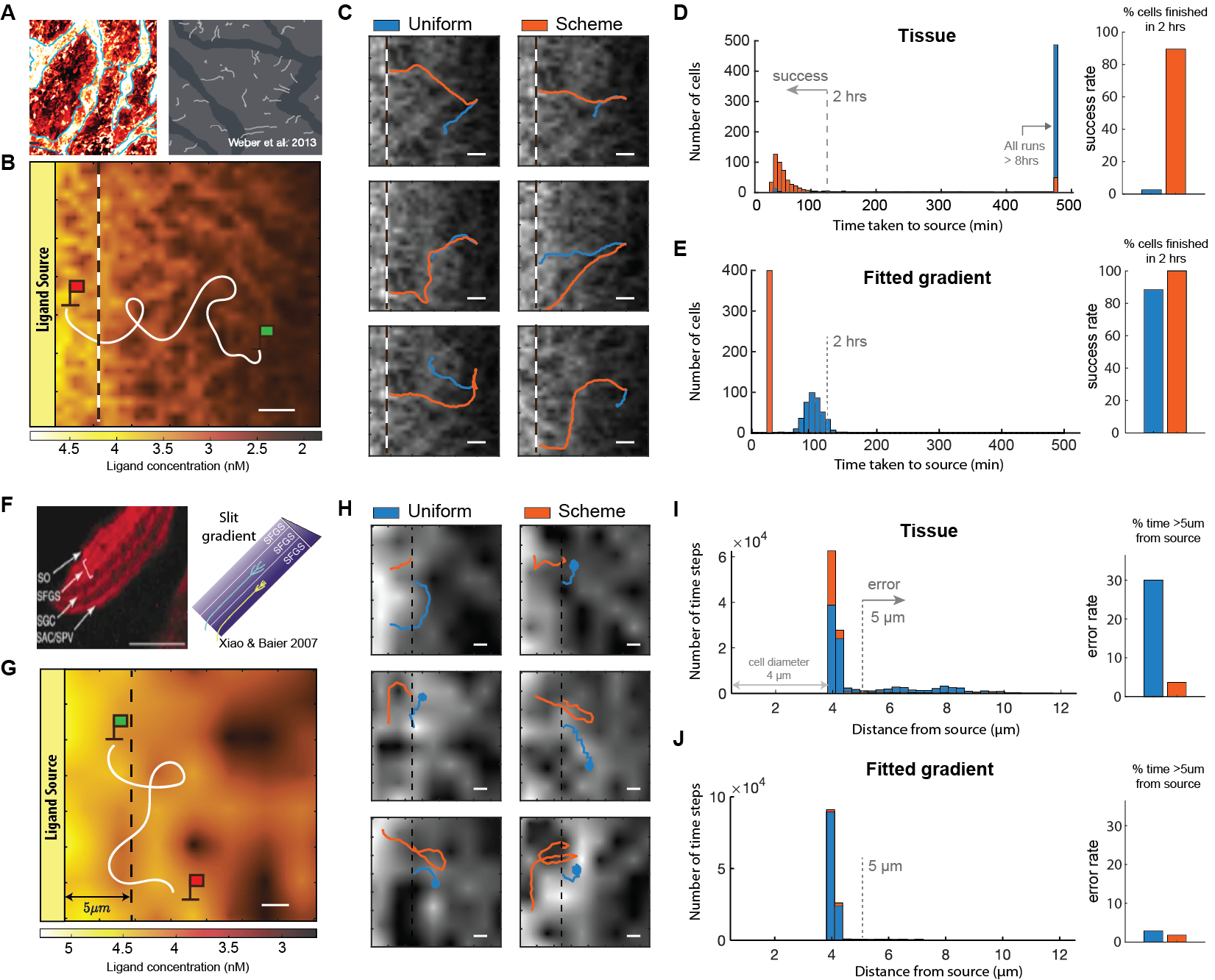}
\centering
\caption{
\textbf{In simulated interstitial gradient, cells localize to source quickly and precisely when receptors are redistributed by the feedback scheme instead of uniformly distributed.}
(A), (left) interstitial CCL21 gradient, (right) white curves represent trajectories of dendritic cells \citep{weber2013interstitial}.
(B), schematic of a navigation task where a cell (green flag) in an interstitial gradient move towards the source (red flag) by sensing spatially-distributed ligands and decoding source direction, scale bar: $10\mu m$.
(C), sample trajectories of repeated simulations of cells ($5\mu m$ radius) navigating with uniform receptors and with scheme-driven receptors, all scale bars: $10\mu m$.
(D), (left) histogram of time taken to reach source across $500$ starting positions of equal distance from source, note the rightmost (blue) bar includes all cells that did not reach the source after $8$ hours; (right) bar plot showing percentage of runs completed in $2$ hrs (success rate).
(E), same type of data as in panel D for cells navigating in an exponential gradient (fitted to the interstitial gradient used to generate panel D).
(F), red stripes (left) represent growth cones moving within specific lamina along a Slit gradient (right schematic), scale bar: $40$ $\mu m$ \citep{xiao2007lamina}.
(G), schematic of a navigation task where a cell (green flag) senses its environment in order to remain close to source, solid white line represent parts of cell trajectory outside a $5$ $\mu m$ lamina, dotted white line represent the parts within the lamina, scale bar: $2$ $\mu m$.
(H), sample trajectories of repeated simulations of cells performing task with either uniform or scheme-driven receptors, all scale bars: $2$ $\mu m$.
(I), (left) histogram of time spent by cell at various distance from the ligand source (measured from source to farthest point on cell, perpendicular to source edge) aggregated across $500$ starting positions along the ligand source; (right) bar plot showing percentage of time spent more than $5$ $\mu m$ from source (error rate).
(J), same type of data as in panel I for cells navigating in an exponential gradient (fitted to the interstitial gradient used to generate panel I). }
\label{fig:haptotaxis}
\end{figure}

	As an application of the receptor placement strategy, we show that cells using our feedback scheme efficiently localizes to the source of simulated interstitial gradients.
	Immune cells can navigate towards the source of an interstitial gradient in a directed, efficient manner (\autoref{fig:haptotaxis}A) \citep{weber2013interstitial}.
	Efficient navigation can be difficult in uneven tissue environments, partly due to the existence of local maxima away from the ligand source, trapping cells on their way to the source (\autoref{fig:haptotaxis}B). 
	By simulating cell navigation using standard models of directional decoding (see Methods), we found that cells with uniform receptors can indeed become trapped during navigation.
	\autoref{fig:haptotaxis}C demonstrates this behavior through the trajectories of individual cells with uniform receptors (blue), as they consistently become stuck within specific regions.
	On the other hand, using the same method of directional decoding, cells with scheme-driven receptors (orange) consistently reach the source in a directed manner.
	\autoref{fig:haptotaxis}D illustrates this difference through a histogram of the time it took for a cell to reach the source, created by simulating $500 \times 2$ cells starting at uniformly-sampled locations $50 \, \mu m$ from the source, moving at a constant speed of $1\mu m/$min.
	Only $2.8\%$ of cells (14/500) with uniform receptors reached the source within $2$ hours, whereas $87\%$ of cells (437/500) using the feedback scheme did so in the same time frame, boosting success rate by $3000\%$.
	In fact, \autoref{fig:haptotaxis}D shows that more than $97\%$ cells with uniform receptors fail to reach the source even after $8$ hours, as expected due to local trapping. 
	This $30$-fold difference in success rate persists for a wide range of scheme parameter values, and for different methods of directional decoding (SI). 
	We emphasize that the poor performance of cells with uniform receptors is not due to inaccuracy associated with decoding local gradients. Indeed, cells that can perfectly decode local gradient show only minor improvements in performance (SI). It is the presence of local concentration peaks that makes canonical gradient sensing with uniform receptors highly ineffective.
	In addition, we found that signal amplification downstream of receptor binding, through a local-excitation, global-inhibition (LEGI) network, did not significantly improve navigation performance.
	Incorporating "memory" to enable temporal averaging improves success rate to around $12\%$, when averaging the past $20$ steps (equivalent of $10$ minutes) (see SI).
	Interestingly, \autoref{fig:haptotaxis}E shows that the difference in performance between uniform and scheme-driven receptors is significantly reduced in an exponential gradient (fitted to the interstitial gradient), a mere $15\%$ difference between cells with uniform vs. scheme-driven receptors  ($87\%$ vs. $100\%$). We discuss the analogy between our feedback scheme and the "Infotaxis" algorithm \citep{vergassola2007infotaxis} in the Discussion section .
	
	As a second application, we assess whether our feedback scheme can help cells retain within a precise region along a chemical gradient.
	During certain developmental programs, cells must restrict their movements within a region along a gradient in order to form stable anatomical structures. 
	Growth cones demonstrate an extraordinary ability in accomplishing this task. Axon projections of retinal ganglion cells can remain within a bands of tissue (lamina) of only $3-7 \mu m$ wide, at a specific point along a chemical gradient (\autoref{fig:haptotaxis}F) \citep{xiao2007lamina}.
	\autoref{fig:haptotaxis}G illustrates how we assess our scheme's ability to achieve this level of precision. We initiate a cell at a gradient source and track the proportion of time the cell was more than $5 \, \mu m$ away from the source. As the cell moves along the gradient, uneven ligand distribution in the environment can lead the cell to move erroneously away from the source.
	\autoref{fig:haptotaxis}H shows that cells with uniform receptors (blue) can indeed make excursions away from the source. Cells with the feedback scheme (orange), however, can reliably stay close to the source for an extended period of time.
	We quantify this difference by pooling from $400$ trajectories of cells starting at different positions along the source, decoding source direction and navigating for $2$ hrs ($240$ time steps). 
	\autoref{fig:haptotaxis}I shows the number of time steps the cells collectively spent at specific distances from the source.
	Considering the $5\, \mu m $ lamina (band) in \autoref{fig:haptotaxis}F, cells with uniform receptors are found outside this lamina $30\%$ of the time ($28464/96000$ time steps). 
	On the other hand, cells with the feedback scheme are outside this lamina $4\%$ of the time ($3965/96000$ steps), a $650\%$ reduction in error rate.
	This difference in error rate persists for a wide range of scheme parameter values, and for different methods of directional decoding (see SI). 
	When the same comparison is made in the corresponding fitted exponential gradient, the difference in performance is considerably smaller. \autoref{fig:haptotaxis}J shows the same error rate is reduced by a mere $50\%$ from cells with uniform to scheme-driven receptors ($3\%$ vs. $2\%$). Taken together, our feedback scheme is functionally effective in (simulated) patchy environments found in tissue, enabling cells to solve common navigation tasks with significantly improved accuracy and precision.
	
\subsection*{Optimal efficacy accurately predicts experimental observations of membrane receptor distribution}

\begin{figure}[H]
\includegraphics[width=15cm]{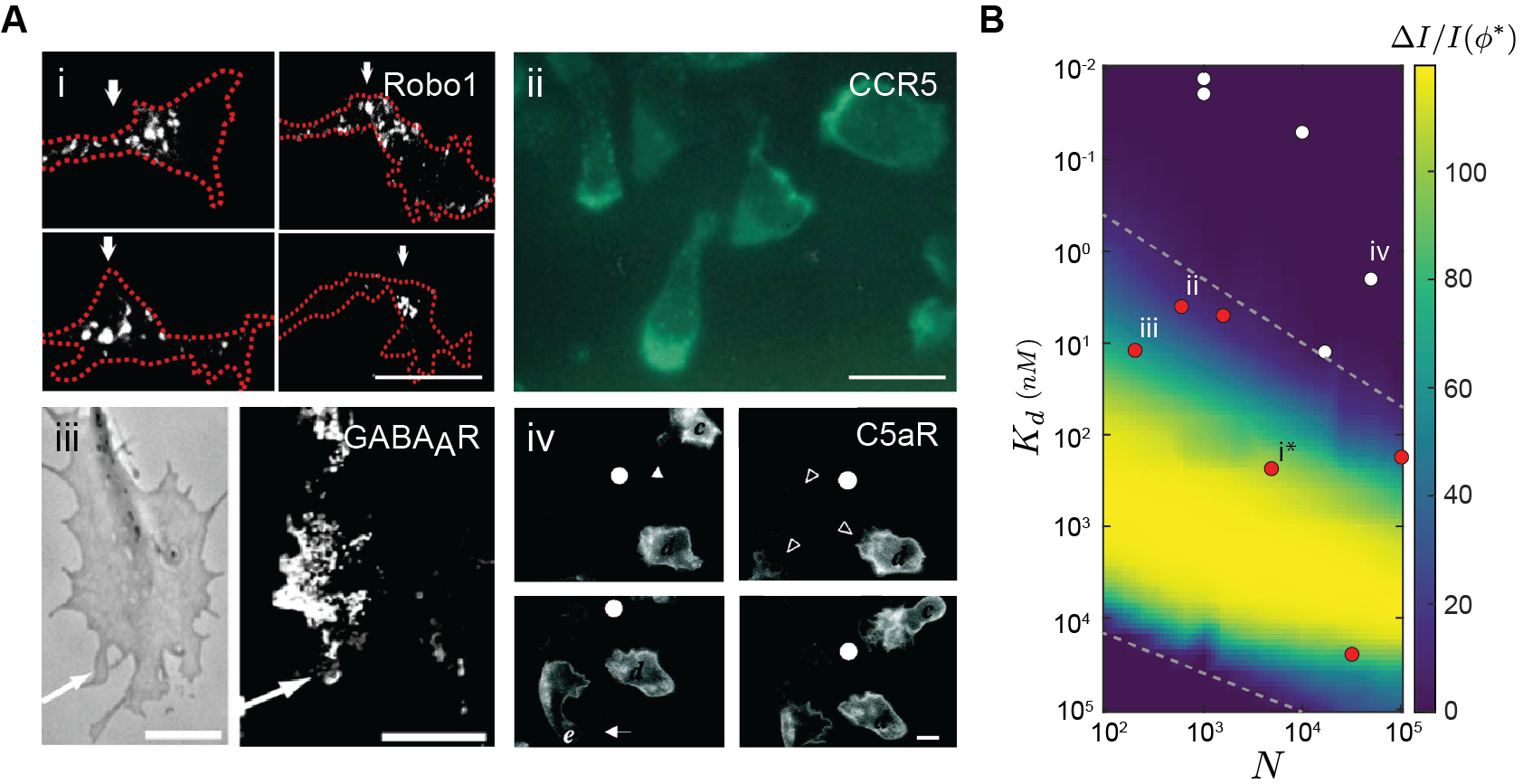}
\centering
\caption{\textbf{Optimal efficacy $\Delta I/I(\phi^*)$ predicts observed distributions of cell surface receptors using their surface expression level and binding affinity.} (A), observed membrane distributions of receptors in heterogeneous environments, i. white arrowheads indicate Slit receptor Robo1 of commissural growth cones navigating in an interstitial Slit gradient \citep{pignata2019spatiotemporal}, ii. chemokine receptor CCR5 of human T lymphocytes subject to a CCL5 gradient \citep{nieto1997polarization}, iii. (left) transmission image of growth cone, white arrowhead indicates direction of GABA gradient, (right) bright dots represent $\text{GABA}_\text{A}$R redistributing in response to a GABA gradient \citep{bouzigues2007asymmetric}, iv. C5aR-GFP remains uniformly distributed in response to a point source of a C5aR agonist, delivered by micropipette (white dot), open arrowheads point to leading edges of cells \citep{servant1999dynamics}. Scale bars i-iii: $5\mu m$, iv: $10\mu m$. (B), optimal relative efficacy $\Delta I/I(\phi^*)$ for different values of $K_d$ and $N$, dashed gray line demarcate region where $\Delta I/I(\phi^*) > 10$; fixed constitutive activity $\alpha = 0.1$; red dots correspond to receptors that polarize in heterogeneous environments, white dots represent receptors that are constantly uniform, roman numerals correspond to receptors in panel A. *Note Robo1 receptor number ($N$) is taken from cancerous non-neuronal cells (see Methods for data taken from literature).}
\label{fig:realcell}
\end{figure}

The optimal efficacy $\Delta I/I(\phi^*)$ accurately predicts experimental observations of receptor distribution over the cell surface.
Recall that in addition to being dependent on the environment, the optimal strategy $\phi^*$ takes as parameters two receptor properties: surface expression level ($N$) and binding affinity ($K_d$).
As a result, optimizing the placement of different types of receptors can lead to different efficacy.
In \autoref{fig:realcell}B, we see efficacy drops rapidly when the ratio of receptor level and binding affinity is either too large or too small. 
Specifically, this result predicts that optimizing receptor placement improves spatial sensing only when $ 10^{-2} \lesssim N/K_d \lesssim 10^{3}$ (dashed lines).
\autoref{fig:realcell}B was generated assuming a tissue environment, where the ratio between average ligand concentration and $K_d$ is fixed.
We test this prediction on real receptors that function in human tissues.
\autoref{fig:realcell}A shows certain receptors (i-iii) indeed polarize toward maximum ligand concentration \citep{pignata2019spatiotemporal, nieto1997polarization, bouzigues2007asymmetric}, in agreement with $\phi^*$, while others (iv) are always uniform \citep{servant1999dynamics}.
As predicted, \autoref{fig:realcell}B shows that receptors known to polarize in non-uniform environments (red) have parameter values corresponding to large $\Delta I/I(\phi^*)$, whereas receptors that are always uniformly distributed (white) have parameter values corresponding to small $\Delta I/I(\phi^*)$ (see Methods). 
This agreement between theory and observations is not meant to imply that evolution optimizes receptor placement. 
Indeed, there are key caveats such as variations in receptor expression over time and differences between the environments of different receptors.
Our theory does, however, provide a framework for studying natural variations in the spatial organization of receptors, such as differences observed between chemotactic receptors in the same T-cell \citep{nieto1997polarization}.

	\section*{Discussion}
	
	\subsection*{Engineering cell to function in natural environments}
	In this work, we explored the question of whether the placement of receptor on the cell surface can be optimized to improve information acquisition, by cells in different environments. We found the optimal strategy involves receptor polarization, and is more effective than uniformly-placed receptors in soils and tissues but not simple, monotonic gradients. Furthermore, a simple feedback scheme implements the strategy and enables efficient navigation simulated along interstitial gradients.
	Our result demonstrates that effective design principles of cell systems can be derived by careful considerations of the spatial structure of the environment in which they function.
	This aspect is especially relevant given rapidly growing interest in deploying engineered cells in natural milieu, as well as recent progress in characterizing natural microenvironments of tissues, soils, and oceans. 
	It will be interesting to extend this general approach to other cell functions \citep{sivak2014environmental}, such as how strategies for cell-to-cell communication depend on spatial structures of the environment.
	
	\subsection*{Adapting framework to optimize other cell properties with respect to environmental statistic}
	
	One can easily adapt our framework to understand how variables other than receptor placement affects spatial sensing. Although this work is about optimizing receptors placement, the key quantity being tuned is the spatial distribution of receptor activity, hence our result is relevant to any variable that 1) affects receptor activity and 2) redistributes across space. To illustrate, consider a generalized model of mean receptor activity,
	\begin{equation}
		\mathbb{E}[A_i \mid c_i] = f(r_i,\bm{\theta}) \Big(\frac{c_i}{c_i + k_D} + \alpha \Big),
	\end{equation}
	
	where $f$ is an unspecified function representing the "effective" number of receptors, and $\bm{\theta}$ represents an arbitrary set of parameters. In this work, we considered the case where $f(r_i,\bm{\theta}) = r_i$ (equation \eqref{appendixchannel2}). In addition to receptor number, other factors such as phosphorylation and membrane curvature also affect local receptor activity $A_i$ by tuning receptor "sensitivity".  In this way, one can optimize spatial sensing by tuning variables other than receptor placement, by choosing the appropriate form for $f$. For example, it is known that given uniformly distributed receptors, those found in regions of higher curvature can exhibit higher activity \citep{rangamani2013decoding}. Suppose we want to know the optimal way to adjust cell shape to maximize information acquisition, given a linear relationship between curvature $\beta$, and "effective" receptor number, i.e. $f(r_i) = \beta r_i$. Assuming uniformly placed receptors and a constraint on total membrane curvature, we quickly arrive at the optimal solution since this problem is now identical to our original formulation. The optimal strategy is to increase curvature at regions of high ligand concentration, by making local protrusions. 
	
	\subsection*{Optimizing spatial organization at different stages of information processing}
	Optimizing information transmission by organizing effectors in space can happen at all stages of signal processing within the cell, but is likely most effective at the receptor level. The most obvious reason is due to the data processing inequality, which states that post-processing cannot increase information. Therefore, only optimization at the level of receptor activation can increase the total amount of information that is available to the cell. The second reason is due to the "hourglass" topology of cell signaling networks, which represent the fact that a large number of signaling inputs converge onto a small number of effectors internal to the cell. For example, GPCRs, one of the largest group of cell surface receptors, drive downstream signaling through the same G-proteins. This feature makes optimizing spatial organization at later stages of information processing very difficult, since information can be easily lost by diffusion of effector molecules activated by different inputs, which ends up "mixing" different spatial signals.
	
\subsection*{Nonlinearity of Poisson channel}
	
Strong polarization of optimal receptor placements can be partly explained by nonlinearity of the (Poisson) channel mutual information with respect to the expectation value of the input. We illustrate this fact by considering the canonical scalar Poisson channel,
	\begin{equation}
		Y|X \sim \text{Pois}(\alpha X + \lambda)
	\end{equation}
	where $X$ is a scalar input analogous to ligand concentration, and $Y$ is a scalar output analogous to receptor activity. The scaling variable $\alpha$ plays a similar role as receptor number and the dark current $\lambda$ can be taken to represent constitutive receptor activity.
	When there is no constitutive activity ($\lambda = 0$) and receptor number is small, we can approximate the derivative of the mutual information with respect to the receptor number,
	\begin{equation}
		\frac{d I(X;Y)}{d\alpha} \approx E(X\log X) - E(X)\log E(X) 
	\end{equation}
	This derivative represents the information content per receptor, and depends only on the input $X$. Intuitively, one should allocate more receptors to channels with larger $\frac{d I(X;Y)}{d\alpha}$ in order to maximize total information. Furthermore, the larger the difference is in $\frac{d I(X;Y)}{d\alpha}$ between two channels, the larger the asymmetry in receptor allocation should be. By considering $X$ as a Poisson random variable and plotting $\frac{d I(X;Y)}{d\alpha}$ as a function of $E(X)$ (see SI), we see the slope of this function is maximized when $E(X)$ is small ($\approx 1$ molecules/$\mu m^2$) and approaches zero as $E(X)$ increases. This nonlinearity suggests that only when ligand concentration is small will a small difference in concentration lead to relatively large difference in information content, leading to strong asymmetry (localization) in receptor allocation.
	Lastly, this asymmetry can be made even stronger for receptors exhibiting constitutive activity ($\lambda > 0$), a property that many receptors including GPCRs have \citep{slack2012development}. 
	
		\subsection*{Connection between information acquisition and navigation}
	A receptor placement strategy aimed at maximizing information rate can  boost cell navigation performance. Since information content increases towards the ligand source, receptors are more likely to move towards the side of the membrane closer to the source rather than away, enforcing movement up gradients. Furthermore, the trade-off between information acquisition and receptor redistribution can be viewed as combining exploitative  and  exploratory tendencies, where larger redistribution "cost" favors exploitation. This is similar in principle to the "infotaxis" algorithm \citep{vergassola2007infotaxis}, where one can view receptors as "navigating agents", whose movements guide the cell towards the target. Although the idea is quite intuitive, the exact relationship between navigation and information acquisition requires further investigation. For instance, the feedback scheme is most effective in the case of limited input sampling (see SI), which suggests maximizing information content per sample indeed helps with navigation.  On the other hand, redistributing receptors to maximize absolute information does not allow cells to navigate much better than uniformly distributed receptors (see SI). Taken together, maximizing information rate appears to be more effective for the particular task of navigation, compared to maximizing absolute information.

	\section*{Acknowledgement}
We thank Michael Elowitz and Erik Winfree for scientific discussions and Dominik Schildknecht, Tae Han Kim, Pranav Bhamidipati,  Abdullah Farooq, for feedback on the manuscript. We also would like to thank Eugenio Marco and Katarzyna Rejniak for technical advice with receptor feedback and tissue simulations, respectively. The authors would like to acknowledge the Heritage Medical Research Institute and Packard Foundation for funding and intellectual support.
	
	\vspace{20pt}
	
	\bibliography{bibliography}
	
\end{document}